
\def\doeack{\foot{Work supported, in part, by the Department of Energy.}}
\def\gev{~{\rm GeV}}
\def\tev{~{\rm TeV}}

\def\fbi{~{\rm fb}^{-1}}


\def\prdj#1{{\it Phys. Rev.} {\bf D{#1}}}
\def\npbj#1{{\it Nucl. Phys.} {\bf B{#1}}}

\def\plbj#1{{\it Phys. Lett.} {\bf B{#1}}}

\def\prepj#1{{\it Phys. Rep.} {\bf {#1}}}

\def\mt{m_t}
\def\wp{W^+}
\def\wm{W^-}
\def\rta{\rightarrow}
\def\tanb{\tan\beta}
\def\sinb{\sin\beta}
\def\cosb{\cos\beta}
\def\cotb{\cot\beta}
\def\lplm{l^+l^-}
\def\cale{{\cal E}}
\def\calo{{\cal O}}

\def\hn{\phi}
\def\hsm{\phi^0}
\def\mhn{m_\hn}
\def\nsd{N_{SD}}
\def\dg{\Delta g}

\def\mz{m_Z}
\def\anti{\overline}

\def\ifmath#1{\relax\ifmmode #1\else $#1$\fi}

\def\3quarter{{\textstyle{3 \over 4}}}

\def\eps{\epsilon}

\input phyzzx
\Pubnum={$\caps UCD-93-2$\cr}
\date{January, 1993}

\titlepage
\vskip 0.75in
\baselineskip 0pt
\hsize=6.5in
\vsize=8.5in
\centerline{{\bf GLUON FUSION: A PROBE OF HIGGS SECTOR CP VIOLATION}\doeack}
\vskip .075in
\centerline{John F. Gunion and  T.C. Yuan}
\vskip .075in
\centerline{\it Davis Institute for High Energy Physics}
\centerline{\it Department of Physics, U.C. Davis, Davis CA 95616}
\vskip .075in
\centerline{and}
\vskip .075in
\centerline{B. Grz\c{a}dkowski}
\vskip .075in
\centerline{\it CERN, Geneva, Switzerland}
\centerline{\it and}
\centerline{\it Institute for Theoretical Physics}
\centerline{\it University of Warsaw, Warsaw, Poland}
\vskip .075in
\centerline{ABSTRACT}
\vskip .075in
\centerline{\Tenpoint\baselineskip=12pt
\vbox{\hsize=12.4cm
\noindent We demonstrate that
CP violation in the Higgs sector, \eg\ of a multi-doublet
model, can be directly probed using gluon-gluon collisions at the SSC.}}
\vskip .15in

Understanding the Higgs sector is one
of the fundamental missions of future high energy colliders such
as the SSC and LHC. In particular, it will be important to
know if CP violation is present in the Higgs sector. Generally,
either spontaneous or explicit CP violation can be present
if the Higgs sector consists of
more than the single doublet field of the Standard Model (SM).
\REF\hhg{J.F. Gunion, H.E. Haber, G. Kane and S. Dawson,
{\it The Higgs Hunter's Guide},
Frontiers in Physics Lecture Note Series \#80,
(Addison-Wesley Publishing Company, Redwood City, CA, 1990).}
(For a review of this and other issues summarized below, see
Ref.~[\hhg], and references therein.)
However, important classes of models with extended Higgs sectors
either do not allow for Higgs sector CP violation or are
inconsistent with current experiment if significant CP violation in the
Higgs sector is present. Among such models, supersymmetric theories
are the most important example. There, a phase for a Higgs field vacuum
expectation value in excess of about $10^{-2}$ would imply imaginary
components for slepton, squark, chargino and neutralino propagators that
would result in electric dipole moments of the electron and neutron
in excess of experimental limits.
Thus, once a Higgs boson is discovered, it will be crucial to determine
whether or not it is a pure CP eigenstate.

Although there are a variety of experimental observables that are indirectly
sensitive to CP violation in the Higgs sector (such as EDM's, top quark
production and decay distributions, \etc), CP-violating contributions
typically first appear at one-loop, or are otherwise suppressed, and
will be very difficult to detect in a realistic
experimental environment.  In addition,
if CP violation in this class of observables is detected, it could easily
arise from sources other than the Higgs sector.  In this letter,
we shall show that the CP nature of a neutral Higgs boson ($\hn$) is directly
probed by the difference between its production rates
through gluon-gluon fusion processes for colliding proton beams of
opposite polarizations.
(The proposed asymmetry is closely analogous to that developed previously for
collisions of polarized back scattered laser beams at a future
linear $\epem$ collider.
\Ref\bslaserbeasm{B. Grz\c{a}dkowski and J.F. Gunion,
\plbj{294} (1992) 361.})
We compute the magnitude of the asymmetry that can be expected at the
SSC in the context of a general two Higgs doublet model (2HDM)
for a variety of models of the polarized gluon distribution function,
$\dg(x)$.  For all but extremely conservative $\dg(x)$ choices,
large asymmetries are possible since the $gg$ coupling to the CP-even
and CP-odd components of the $\hn$ are generically comparable
(both arising at one loop). Indeed,
we find that asymmetries larger than 10\% are quite typical; these would
be observable in the $\hn\rta ZZ\rta \lplm X$ final state after 1-3
years of running.
In the computations quoted here, we consider the situation in which
the only extension of the SM occurs in the Higgs sector --- $\hn$ production
rates and asymmetries are generally larger
in theories containing additional heavy colored fermions.

The procedure for computing the $gg\rta\hn$ cross section
in leading order is well-known.\refmark{\hhg}
Our computations will employ the leading order formalism, but it should
be noted that radiative corrections to this procedure
have been computed, and for a typical value of $\alpha_s$ result in
an enhancement factor of about 1.7.
\Ref\dawzer{ S. Dawson, \npbj{359}
(1991) 283; A. Djouadi, M. Spira, and P. Zerwas, \plbj{264} (1991) 440.}
In this sense, our results will be conservative.

Crucial to our discussion is the degree of polarization that
can be achieved for gluons at the SSC. The amount of gluon polarization
in a positively-polarized proton beam,
defined by the structure function difference $\dg(x)=g_+(x)-g_-(x)$,
is not currently known with any certainty.
(Here, the $\pm$ subscripts indicate gluons with $\pm$ helicity, and
 $g(x)=g_+(x)+g_-(x)$ is the unpolarized gluon distribution function.)
The relative behavior of $\dg(x)$ compared to $g(x)$ is theoretically
constrained in the $x\rta1$ and $x\rta0$ limits:
$\dg(x)/g(x)\rta 1$ for $x\rta 1$ and $\dg(x)/g(x)\propto x$ for $x\rta 0$.
Simple models which satisfy these constraints suggest
\Ref\brodskymodel{See, for example, S. Brodsky and I.A. Schmidt,
\plbj{234} (1990) 144.}
that a significant amount of the proton's spin could be carried by the gluons.
The EMC
\Ref\emc{European Muon Collaboration (J. Ashman, \etal), \npbj{328} (1989) 1.}
data on the polarized structure function $g_1^p(x)$ is also most easily
interpreted if this is the case.
\Ref\rossreview{See, for example, G. Ross, {\it Lepton/Photon Symposium},
Stanford CA, August 7-12 (1989), p. 41.}

We shall employ a variety of models that have appeared in the literature.
\REF\chengwai{H.-Y. Cheng and C.F. Wai, \prdj{46} (1992) 125.}
In one extreme, also considered in Ref.~[\chengwai], we assume that $\dg(x)=0$
at all $x$ when $Q^2=10\gev^2$. $Q^2$ evolution will retain $\dg\equiv\int_0^1
\dg(x)\,dx=0$
(\ie\ gluons never carry any portion of the proton's spin),
but $\dg(x)$ will develop substantial oscillations at the large $Q$
values of interest for Higgs production. Another extreme
is to assume that none of the proton's spin can be carried by strange quarks.
This is the second case considered in Ref.~[\chengwai],
and leads to large $\dg$, $\dg\sim 4.5$ at $Q^2=10\gev^2$.
Aside from numerical differences, this is also the choice
considered in Ref.~[\rossreview].
We shall label this as case (2). We employ the detailed $\dg(x)$
form given in Ref.~[\chengwai]. We also compute results for
an intermediate choice, case (3), of $\dg\sim 2$ (at $Q^2=10\gev^2$)
considered in Ref.~[\chengwai], using their parameterization for $\dg(x)$.
Two additional $\dg(x)$ parameterizations
have also been employed.  These are: the the Berger-Qiu parameterization
\Ref\bergerqiu{E.L. Berger and J. Qiu, \prdj{40} (1989) 778.}
$\dg(x)=g(x)~~(x>x_c)$, $\dg(x)= {x\over x_c}g(x)~~(x<x_c)$,
where $x_c\sim 0.2$ yields a value of $\dg\sim 2.5$ at $Q^2=10\gev^2$,
case (4); and the rather modest $\dg(x)$ proposal of
\REF\bourrely{C. Bourrely, J. Soffer, F.M. Renard and P. Taxil,
\prepj{177} (1989) 319.}
Ref.~[\bourrely], with $\dg\sim 0.2$ at $Q^2=10\gev^2$, case (5).
Quark distributions can be chosen, in association with all the
$\dg(x)$ forms adopted in the above five cases, that reproduce
the normal deep inelastic data and the polarized proton EMC data.
In obtaining results for Higgs production,
we have computed the evolved $\dg(x)$
starting with the $Q^2=10\gev^2$ inputs specified in cases (1--5),
using standard polarized structure function evolution.
\Ref\tung{The program employed is that developed by W.-K. Tung.}

The asymmetry we compute is simply
$A\equiv[\sigma_+-\sigma_-]/[\sigma_++\sigma_-]$,
where $\sigma_{\pm}$ is the cross section for Higgs production in collisions
of an unpolarized proton with a proton of helicity $\pm$, respectively.
$\sigma_+-\sigma_-$ is proportional to the integral over $x_1$ and $x_2$
(with $x_1x_2=m_\phi^2/s$) of
$g(x_1)\dg(x_2)\left[|{\cal M}_{++}|^2-|{\cal M}_{--}|^2\right]$,
while $\sigma_++\sigma_-$ is determined by the integral of
$g(x_1)g(x_2)\left[|{\cal M}_{++}|^2+|{\cal M}_{--}|^2\right]$.
(We have assumed that it is proton 2 that is polarized.
Distribution functions will be evaluated at $Q=\mhn$.) Now,
$|{\cal M}_{++}|^2-|{\cal M}_{--}|^2$ vanishes for
a CP eigenstate, but can be quite large in a general 2HDM.  We find
$|{\cal M}_{++}|^2-|{\cal M}_{--}|^2\propto -4{\rm Im\,}(\cale\calo^*)$
and
$|{\cal M}_{++}|^2+|{\cal M}_{--}|^2\propto 2\left(|\cale|^2+|\calo|^2\right)$,
where $\cale$ ($\calo$) represents the
$gg$ coupling to the CP-even (-odd) component of $\hn$.
These depend upon the reduced CP-even (scalar, $s$) and
CP-odd (pseudoscalar, $p$) couplings given by
\foot{Reduced couplings are defined relative to SM-like couplings.}
$s_{t\anti t}={u_2\over\sinb}$, $p_{t\anti t}=-u_3\cotb$,
$s_{b\anti b}={u_1\over\cosb}$, and $p_{b\anti b}=-u_3\tanb$.
\REF\weinii{S. Weinberg, \prdj{42} (1990) 860.}
\REF\topas{B. Grz\c{a}dkowski and J.F. Gunion,  \plbj{287} (1992) 237.}
Here, the $u_i$ specify the eigenstate $\hn$ in
the $\Phi_i$ basis of Ref.~[\weinii] (see Ref.~[\topas] for more details).
In a 2HDM, $\sum_i u_i^2=1$, but they are otherwise unconstrained.
Results for the SM Higgs boson correspond to
taking $u_1=\cosb$, $u_2=\sinb$, and $u_3=0$. More generally,
for a CP-even eigenstate
we would have $u_3=0$, while for a CP-odd eigenstate $|u_3|=1$.
We note that the widths for the $\hn$ to decay to
$b\anti b$ and $t\anti t$ are determined
using these reduced couplings by appropriately
weighting the results for CP-even and CP-odd scalars as given in
Appendix B of Ref.~[\hhg]. $ZZ$ and $\wp\wm$ widths are obtained
using the reduced scalar coupling $s_{\wp\wm,ZZ}=u_2\sinb+u_1\cosb$.

\FIG\ggcphiggsnsd{}
\midinsert
\vbox{\phantom{0}\vskip 4.5in
\phantom{0}
\vskip .5in
\hskip -20pt
\special{ insert scr:ggcphiggs_nsd.ps}
\vskip -1.4in }
\centerline{\vbox{\hsize=12.4cm
\Tenpoint
\baselineskip=12pt
\noindent
Figure~\ggcphiggsnsd:
Maximal statistical significance, $\nsd^{max}$,
achieved for the asymmetry signal in the $\phi\rta ZZ\rta\lplm X$ channel as
a function of $m_\phi$ at the SSC with $L=10\fbi$.
The curves for different $\dg(x)$ choices are labelled by
the case number, 1--5.
}}
\endinsert

To obtain a numerical indication of the observability of $A$, we have
proceeded as follows. We assume that $\phi$ can be best
detected in the $\phi\rta ZZ\rta \lplm X$ modes (where
$l=e,\mu$ and we include all possible $X$ ---
$X=\lplm,\tau\anti\tau,q\anti q,\nu\anti \nu$ ---
so that the net branching ratio for $ZZ\rta\lplm X$ is $\sim0.134$).
We compute the statistical significance of the asymmetry signal as
$\nsd\equiv (N_+-N_-)/\sqrt{N_++N_-}$, where $N_+$ ($N_-$) is the
number of events predicted for positive (negative) proton polarization
in the $ZZ\rta\lplm X$ mode. Since $\dg(x)\rta g(x)$ at large $x$,
we impose a cut on the Higgs boson
events designed to enhance the importance of large $x_2$
in the convolution integrals contributing to the numerator and
denominator of the asymmetry $A$.
The appropriate cut takes the form $x_F^{\hn}=x_1-x_2<x_F^{cut}$.
For each value of $\mhn$ and each $\dg(x)$ case we search
for the choice of $x_F^{cut}$ which optimizes $\nsd$; this optimal $x_F^{cut}$
is independent of the Higgs sector CP violation parameters.
Finally, we search (at fixed $\tanb=v_2/v_1$) for the
parameters of the most general CP-violating 2HDM that yield
the largest achievable statistical significance, $\nsd^{max}$.
Of course, it will be noted that our estimate for $\nsd$ does not include
the $ZZ$ continuum background, other $\hn$ production
mechanisms, the amount of polarization that can
be actually achieved at the SSC, nor other possible channels in which
the $\phi$ could be detected. We shall comment on these and other issues
shortly.

\FIG\ggcphiggsasym{}
\midinsert
\vbox{\phantom{0}\vskip 4.5in
\phantom{0}
\vskip .5in
\hskip -20pt
\special{ insert scr:ggcphiggs_asym.ps}
\vskip -1.4in }
\centerline{\vbox{\hsize=12.4cm
\Tenpoint
\baselineskip=12pt
\noindent
Figure~\ggcphiggsasym:
Fractional asymmetry, $A$, for which $\nsd$ is maximal, as
a function of $m_\phi$ at the SSC with $L=10\fbi$.
The curves for different $\dg(x)$ choices are labelled by
the case number, 1--5.
}}
\endinsert

The results for $\nsd^{max}$ at the SSC with integrated
luminosity of $10\fbi$ appear in Fig. \ggcphiggsnsd, for $\tanb=2$ and $10$,
and $\mt=150\gev$.  Detection of this asymmetry is clearly not out
of the question.  The reason that significant $\nsd$ values can be achieved
becomes clear from the plot of the corresponding values of $A$,
Fig.~\ggcphiggsasym.  Quite large $A$ values are achieved in the
more favorable $\dg(x)$ models (2) and (4). It should be noted that
the Higgs sector parameters required
to achieve the illustrated $\nsd^{max}$ results are not at all fine tuned.
Large ranges of parameter space yield values very nearly as big.
The large difference between the $\nsd^{max}$ results in cases
(1) and (5), illustrated in Fig.~\ggcphiggsnsd, despite the
close similarity in the $A$ values, Fig.~\ggcphiggsasym,
is due to the much smaller event rates for case (1) compared to
other cases, including (5).  This difference arises because
of the strong $x_F^{cut}$ needed to probe only one sign of the oscillating
$\dg(x)$ of case (1) (thereby allowing for significant $A$).

It is amusing to note that, without a determination of $A$,
the $\hn$ is not necessarily so easily distinguished from
a SM Higgs boson ($\hsm$) of the same mass. For instance,
for the parameter choices which yield $\nsd^{max}$,
both the $\hn$ and $\hsm$ total production rates
and the $\hn\rta ZZ$ and $\hsm\rta ZZ$ branching ratios are similar.
Of course, the total width
of the $\hn$ is generally somewhat smaller than that of the $\hsm$ since
the dominant $\wp\wm$ and $ZZ$ widths are suppressed.
However, the resolution needed to distinguish
the $\hn$ from the $\hsm$ is unlikely to be adequate for $\mhn\lsim 400\gev$.

Our ability to detect $A$ may be either better or worse than that illustrated
in Fig. \ggcphiggsnsd. If only partial polarization, $P$, for
the proton beam can be achieved $\nsd^{max}\rta P\nsd^{max}$. Expectations
\Ref\sylee{S.Y. Lee, private communication.} are that $P$ of about 0.7
can be achieved at the SSC with the introduction of appropriate
siberian snakes \etc\ into the injector and main rings of the SSC.
Limited acceptance efficiency, $\epsilon$, for the final states of
interest yields $\nsd^{max}\rta \sqrt{\epsilon}\nsd^{max}$.

As noted earlier, in computing $\nsd^{max}$ in the $ZZ$ channel, we have not
accounted for the $ZZ$ continuum background.  If this background is large,
it would significantly dilute $A$ since it would yield an additional
contribution to the $N_++N_-$ denominator of $A$, and negligible contribution
to the $N_+-N_-$ numerator. Since the $\hn$ is distinctly narrower than
the SM $\hsm$, this contamination is not so large as one might guess.
Below we shall compute the effect of the $ZZ\rta\lplm X$
continuum background upon the observability of $A$.

Similarly, $WW$ fusion production of the $\hn$
would not contribute significantly
to $N_+-N_-$, but would add to $N_++N_-$. We have estimated
its effects and found them to be insignificant
(at $m_t=150\gev$) for Higgs masses below $800\gev$.
For $\mhn$ between $800\gev$ and $1\tev$,
$\nsd^{max}$ is reduced by at most 15\% due to dilution from $WW$ fusion.
For this high mass region, it might prove beneficial to veto against
the energetic spectator jets at high rapidity associated with the
$WW$ fusion mechanism.  Such vetoing can be done with little
affect upon the $gg$ fusion events of interest.

\FIG\ggcphiggsssclum{}
\midinsert
\vbox{\phantom{0}\vskip 4.5in
\phantom{0}
\vskip .5in
\hskip -20pt
\special{ insert scr:ggcphiggs_ssclum.ps}
\vskip -1.4in }
\centerline{\vbox{\hsize=12.4cm
\Tenpoint
\baselineskip=12pt
\noindent
Figure~\ggcphiggsssclum:
We plot the number of $10\fbi$ SSC years required to detect
$A$ at the $\nsd^{max}=5$ sigma level. The different
curves are for the 5 different $\dg(x)$ cases.
Fractional polarization of $P=0.7$ is employed.
The $ZZ\rta\lplm X$ continuum background has been included after
imposing an approximate acceptance cut on both it and the Higgs
signal characterized by $z_0=0.7$ (see text).
}}
\endinsert

In summary, we should combine a polarization fraction of $P\sim0.7$,
a reasonable acceptance efficiency,
and some $ZZ$ continuum dilution in estimating realistically
achievable $\nsd^{max}$ values.  We have done this numerically as
follows.  We have computed the $ZZ$ continuum and the $\hn\rta ZZ$
rates by imposing an angular cut on the outoing $Z$'s in the $ZZ$
center of mass. We require $|z|<z_0=0.7$ (where $z$ is the cosine of
the angle of one of the $Z$'s with respect to the beam direction);
this corresponds to an acceptance of $\eps=0.7$.  For such a cut,
most $ZZ\rta\lplm X$ events will fall within the usable portion
of a typical detector.  The $ZZ$ continuum is integrated over
a mass range given by $\Delta m_{ZZ}={\rm max}\left\{1.5 \Gamma_T(\hn),
10\gev\right\}$. For the most part, the result is that the $\nsd^{max}$ values
plotted in Fig.~\ggcphiggsnsd\ should perhaps be multiplied by
about 0.5 for a conservative estimate of the achievable statistical
significance
for an observation of $A$ in the $ZZ\rta\lplm X$ channel.
In Fig.~\ggcphiggsssclum\ we display the number of SSC $10\fbi$ years
required to achieve $\nsd^{max}=5$ for $z_0=0.7$ and $P=0.7$.
This plot makes it clear that
there is a reasonable chance of observing or placing a meaningful
bound on $A$, if $\dg(x)$ is cooperative,
in 1-10 SSC years, at least for Higgs boson masses above
about $2\mz$ and below about $500-700\gev$. Results for the LHC are similar.
For measuring $A$, a $100\fbi$ LHC year is just slightly better than a
$10\fbi$ SSC year. Of course, it should be kept in mind that
determination of $A$ is certainly a second generation
experiment, and it is quite likely that the SSC could
achieve $100\fbi$ per year by the time this experiment is performed.

It is important to reemphasize the uncertainties
associated with $\dg(x)$.  It is clear that if $\dg(x)$
is typified by our cases (2) or (4), then detection of $A$
could prove to be relatively straightforward.
Given the theoretical constraints on
the $x\rta 0$ and $x\rta 1$ limits of $\dg(x)$, and the models
that have been constructed which incorporate these constraints,
we do not regard such favorable forms of $\dg(x)$ as particularly
unlikely. Certainly, cases (1) and (5) seem to be somewhat extreme
in their conservatism.  In our opinion, case (3) could be employed
as a reasonable lower bound for use in planning. Were this
close to the true $\dg(x)$, then observing or bounding $A$
will generally require running the SSC at enhanced luminosity of order
$100\fbi$ per year.

Finally, it should not be forgotten that all our predictions are
based upon the assumption that the heaviest colored
fermion that acquires its mass via the Higgs mechanism is the top quark.
For $\mhn>2m_t$, the addition of a new generation of quarks
yields a large increase in the observability of $A$
(not to mention the observability of the $\hn$ in the first place).
For $\dg(x)$ case (3), at most 3 SSC years would
be required to measure $A$ in the $2m_t<\mhn\lsim 1\tev$ range.

In conclusion, we emphasize that
the ability to polarize one of the proton beams at the SSC or LHC will provide
a unique opportunity for determining the CP nature of any observed neutral
Higgs
boson.  Indeed, if the Higgs boson has both significant CP-even and CP-odd
components, then a large asymmetry between production rates for
positively versus negatively polarized protons will arise
if a reasonable amount of the proton polarization is transmitted
to the gluon distributions.
If measurable CP violation is found in the Higgs sector
many otherwise very attractive models will be eliminated, including
the Standard Model and most supersymmetric models.
In fact, we have noted that measurement of the polarization asymmetry might
be the only tool that will clearly distinguish a Higgs
boson that is a mixed CP eigenstate from the SM Higgs boson (or a Higgs boson
with SM-like couplings). This should provide a rather strong
motivation for expending the relatively modest monetary amounts
needed to achieve polarized SSC or LHC beams.

\smallskip\noindent{\bf 5. Acknowledgements}
\smallskip
We are grateful to H.-Y. Cheng for helpful conversations
and to C.F. Wai and W.-K. Tung for supplying the
polarized structure function evolution programs employed.
\smallskip
\refout
\end